%% file: main.tex
\documentclass[manuscript,acmsmall,table]{acmart}
\definecolor{islamicgreen}{rgb}{0.0, 0.56, 0.0}

\usepackage[]{url}
\usepackage{listings}
\usepackage{tabularx}
\usepackage{graphicx}
\usepackage{booktabs}
\usepackage{enumitem}
\newlist{researchquestions}{enumerate}{1}
\setlist[researchquestions]{label*=\textbf{RQ\arabic*}}
\usepackage{float}
\usepackage{afterpage}
\usepackage{multirow}
\usepackage{graphicx}
\usepackage{rotating}
\usepackage{makecell}

\usepackage{array}
\usepackage[super]{nth}
\usepackage{fnpct}
\usepackage{caption}
\usepackage{enumitem}
\usepackage{xcolor}

\newlist{questions}{enumerate}{2}
\setlist[questions,1]{label=RQ\arabic*.,ref=RQ\arabic*}
\setlist[questions,2]{label=(\alph*),ref=\thequestionsi(\alph*)}
\definecolor{codegreen}{rgb}{0,0.6,0}
\definecolor{codegray}{rgb}{0.5,0.5,0.5}
\definecolor{codepurple}{rgb}{0.58,0,0.82}
\definecolor{backcolour}{rgb}{0.95,0.95,0.92}

\lstdefinestyle{mystyle}{
    backgroundcolor=\color{backcolour},   
    commentstyle=\color{codegreen},
    keywordstyle=\color{magenta},
    numberstyle=\tiny\color{codegray},
    stringstyle=\color{codepurple},
    basicstyle=\ttfamily\footnotesize,
    breakatwhitespace=false,         
    breaklines=true,                 
    captionpos=b,                    
    keepspaces=true,                 
    numbers=left,                    
    numbersep=5pt,                  
    showspaces=false,                
    showstringspaces=false,
    showtabs=false,                  
    tabsize=2
}

\newcommand{\grayrow}{\rowcolor{gray!10}}

\usepackage{hyperref}

\title{Is ChatGPT a Good Software Librarian? An Exploratory Study on the Use of ChatGPT for Software Library Recommendations}

\begin{document}

\raggedbottom

\author{Jasmine Latendresse}
\affiliation{%
  \institution{Concordia University}
  \city{Montreal}
  \country{Canada}}
\email{jasmine.latendresse@mail.concordia.ca}
\author{SayedHassan Khatoonabadi}
\affiliation{%
  \institution{Concordia University}
  \city{Montreal}
  \country{Canada}}
\email{sayedhassan.khatoonabadi@concordia.ca}
\author{Ahmad Abdellatif}
\affiliation{%
  \institution{University of Calgary}
  \city{Calgary}
  \country{Canada}}
\email{ahmad.abdellatif@ucalgary.ca}
\author{Emad Shihab}
\affiliation{%
  \institution{Concordia University}
  \city{Montreal}
  \country{Canada}
   }
\email{emad.shihab@concordia.ca}




\input{abstract}

\maketitle



\input{introduction}

\input{study_design}

\input{results}

\input{discussion}

\input{related_works}

\input{threats}

\input{conclusion}

\bibliographystyle{IEEEtran}
\bibliography{main}

\end{document}

%% file: abstract.tex
\begin{abstract}

Software libraries play a critical role in the functionality, efficiency, and maintainability of software systems. As developers increasingly rely on Large Language Models (LLMs) to streamline their coding processes, the effectiveness of these models in recommending appropriate libraries becomes crucial yet remains largely unexplored. In this paper, we assess the effectiveness of ChatGPT as a software librarian and identify areas for improvement. We conducted an empirical study using GPT-3.5 Turbo to generate Python code for 10,000 Stack Overflow questions. Our findings show that ChatGPT uses third-party libraries nearly 10\% more often than human developers, favoring widely adopted and well-established options. However, 14.2\% of the recommended libraries had restrictive copyleft licenses, which were not explicitly communicated by ChatGPT. Additionally, 6.5\% of the libraries did not work out of the box, leading to potential developer confusion and wasted time. While ChatGPT can be an effective software librarian, it should be improved by providing more explicit information on maintainability metrics and licensing. We recommend that developers implement rigorous dependency management practices and double-check library licenses before integrating LLM-generated code into their projects.

\end{abstract}

\keywords{Software engineering, software ecosystems, large language models.}

%% file: introduction.tex
\section{Introduction}\label{sec:introduction}

Modern software development thrives on code reuse. Open source libraries provide pre-written code blocks, significantly reducing development time and effort~\cite{basili1996reuse, decan2019empirical, murphy2019predicts}. While libraries enhance functionality, they also introduce dependencies --- interconnections between code components --- that can lead to increased complexity and dependency management challenges~\cite{Latendresse_ASE2022, mujahid2021effective, wang2020empirical}.

One critical aspect of dependency management is library selection~\cite{mujahid2023characteristics}. Choosing the right library impacts factors like code maintainability, performance, and security. Previous studies have explored how developers select libraries, and highlighted primarily ad-hod processes based on past experiences, expert advice, and online resources~\cite{hauge2009empirical, haenni2013categorizing}. 


The current, often informal, process of selecting libraries presents an opportunity for new tools to assist developers. Large Language Models (LLMs) are emerging as powerful programming assistants, capable of generating code and streamlining development workflows~\cite{liang2024large}. They have shown potential in improving developer productivity through functionalities like code completion and search~\cite{ross2023programmer, heitz2024evaluation}, but their reliability as "software librarians" (recommending appropriate libraries) remains unexplored. 

Our work aims to explore a specific aspect of LLM functionality: their potential as software librarians for recommending libraries in code generation tasks. In this paper, we investigate the effectiveness of LLMs in such context by conducting an empirical study using ChatGPT~\cite{openai-chatgpt} (specifically \texttt{GPT 3.5 Turbo}) to generate Python code for a large set of coding problems derived from real-world Stack Overflow~\cite{stackoverflow} questions. This allows us to assess LLM performance in a context that mimics actual developer usage. By analyzing the libraries recommended by ChatGPT, we aim to answer the following research questions: 

\begin{researchquestions}
    \item{} \textbf{What are the characteristics of the software libraries recommended by ChatGPT?} Our results show that ChatGPT uses third-party libraries almost 10\% more than human developers. It also tends to use stable and well-established libraries, with fewer dependencies. Moreover, the licenses of ChatGPT's are mostly permissive, but also include copyleft and public domain options. However, our analysis higlight potential challenges for developers, as 6.5\% of ChatGPT library recommendations resulted in import or installation failures. To shed light on these challenges and understand the specific reasons behind these failures, we delve deeper in our second research question.
    
    \item{} \textbf{What are the challenges encountered by developers when using ChatGPT for library recommendations?} We find that 6.5\% of the libraries used by ChatGPT lead to import or installation failures, mainly because of aliases (alternative names for existing libraries), implicit imports of modules (no mention of the explicit parent library), placeholder suggestions, and context-specific recommendations that might not generalize well. 
\end{researchquestions}

Our findings shed light on how well LLMs recommend libraries and the factors that influence their selection. This will not only inform the development of more robust LLM software librarians but also provide actionable recommendations for practitioners to fully benefit from LLM-based tools in streamlining their workflows. We also identify areas where LLMs excel and areas where developers need to exercise caution. Our work makes the following contributions: 

\begin{itemize}
    \item Empirical evidence on the characteristics of libraries used in GPT-generated code and the potential challenges in using these libraries out of the box.
    \item A set of actionable recommendations for developers using LLMs to streamline their workflows, including considerations for potential issues with LLM-suggested libraries.
    \item A dataset of ChatGPT-generated Python code alongside human-written solutions for Stack Overflow prompts, enabling a direct comparison of library selection practices~\footnote{\url{https://zenodo.org/records/11506864}}.
\end{itemize}

\noindent\textbf{Paper Organization.} The rest of the paper is organized as follows. Section \ref{sec:study_design} describes our dataset and methodology. Section \ref{sec:results} presents the findings of our two research questions. Section \ref{sec:discussion} discusses the implications of our findings along with our recommendations. Section \ref{sec:related_works} discusses related work. Section \ref{sec:threats} outlines the threats to the validity of our study. Section \ref{sec:conclusion} concludes this paper.

%% file: study_design.tex
\section{Study Design}
\label{sec:study_design}

\begin{figure*}
    \centering
    \includegraphics[scale=0.5]{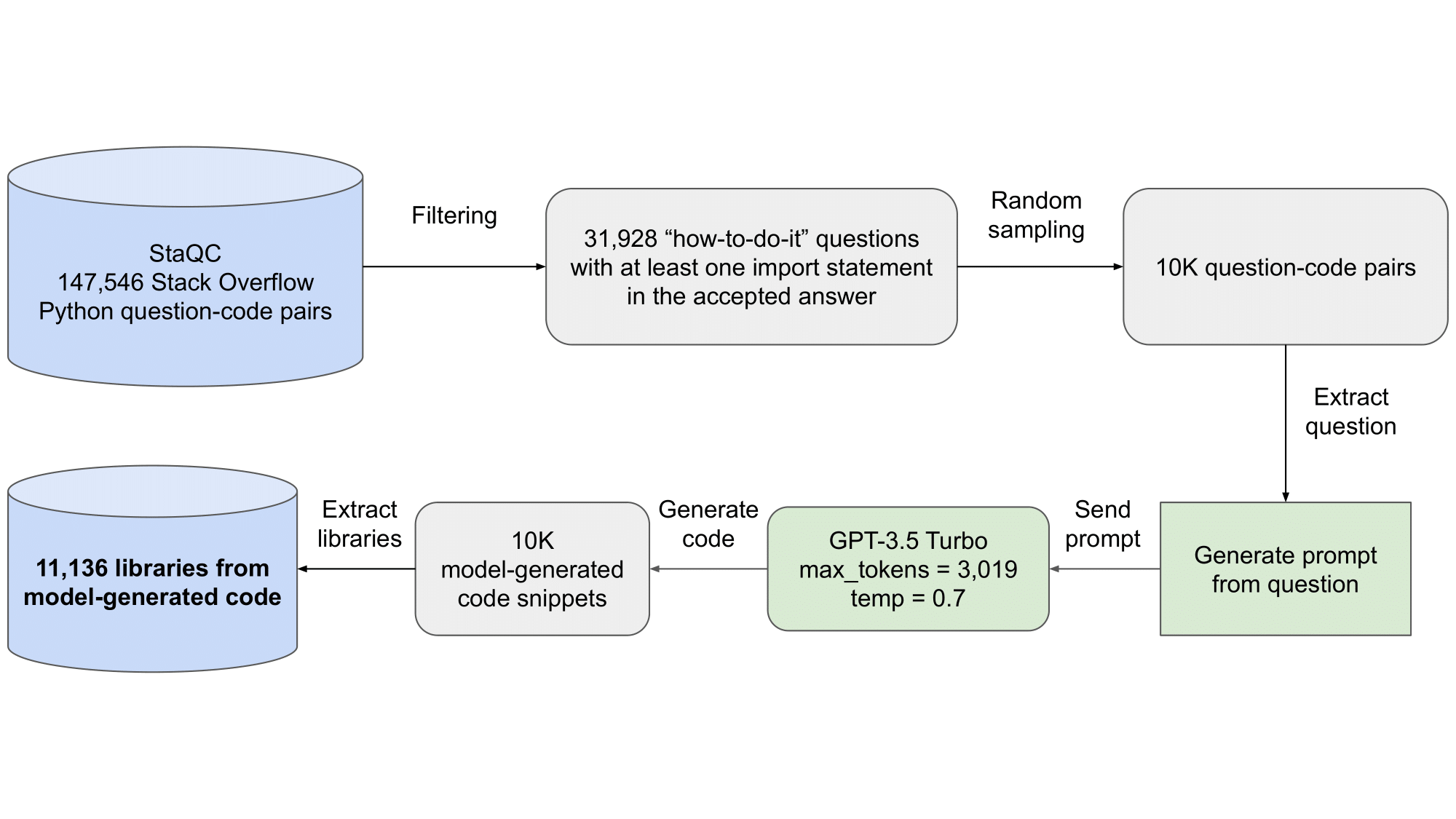}
    \caption{Overview of our approach for collecting libraries ChatGPT-generated code snippets.}
    \label{fig:approach}
\end{figure*}

Figure~\ref{fig:approach} presents an overview of our approach to curate the dataset used in this study. In order to conduct our qualitative and quantitative analysis, we focus on the Python programming language for its popularity, extensive library ecosystem, as well as the demonstrated capabilities of LLMs in generating and understanding Python code~\cite{saabith2020popular, srinath2017python, adamson2023assessing, ellis2024chatgpt}. Below we describe the steps to curate the dataset used in our study and the experimental setup. 

\subsection{Dataset}
\label{sec:dataset}
We prepared our dataset using the dataset of Stack Overflow question-code pairs curated by Yao et al.~\cite{yao2018staqc}. Their set contained 147,546 Python question-code pairs automatically mined from Stack Overflow using a bi-view hierarchical neural network. We selected this dataset because it provides human-written code, which is necessary for our comparative analysis with ChatGPT-generated code. Also, the dataset is comprised of "how-to-do-it" Stack Overflow questions (e.g., how to grab from JSON in selenium python~\footnote{\url{https://stackoverflow.com/questions/26661808/how-to-grab-from-json-in-selenium-python}}), which facilitates prompt crafting for our experiment. Finally, the dataset is labeled, allowing us to filter for specific types of answers, such as multi-code versus single-code answer. Thus, to facilitate code extraction, we select questions annotated single-code answer posts, which are accepted answers that contain only one code snippet~\footnote{\url{https://stackoverflow.com/questions/36479006/how-to-post-data-and-binary-data-using-urllib2-in-python}}. Finally, to ensure that the selected questions are specifically about library usage, we only keep the ones where the code snippet of the corresponding accepted answer includes at least one import statement. This process yields a total of 31,928 question-code pairs. Then, to ensure that our analysis is manageable, we select a random sample of 10,000 question-code pairs (confidence level of 99\% with a $\sim$1\% confidence interval).




\subsection{Experiment Setup}
\label{sec:exp_setup}
To create model-generated code for each Stack Overflow accepted answer, we take a two-step approach. The first step is to construct the prompts to specify the type of tasks we expect the model to complete. In our case, the model will be generating code to answer programming questions in Python. Therefore, the prompts are structured with the base prompt "Generate Python code for the following question" to which we append the Stack Overflow questions. Then, to facilitate the post-processing of responses, we ask the model to "only generate code, no explanations". Thus, the final prompt is "Generate Python code for the following question: <question>. Only generate code. No explanations.".


Then, after generating the prompts, we feed them to the model to generate the corresponding code. To achieve this, we use the OpenAI API to query the language model \texttt{GPT-3.5 Turbo}. We selected this specific model due to its free availability, which lowers barriers to use, enhancing the inclusivity of our study and simplifying replication, as well as for its rapid response time. This corroborates with the work of Destefanis et al.~\cite{destefanis2023preliminary} who have shown that \texttt{GPT-3.5 Turbo} outperforms other models in generating correct code, with a 90.6\% correctness rate across various problem categories. 

The parameters that we used for the model are shown in Figure~\ref{fig:approach}. These parameters are essentially to guide the model's response generation. One such parameter is \textit{temperature} (\texttt{temp}), which affects the creativity or randomness of the response. In our study, we use ChatGPT's default temperature which is of 0.7~\footnote{\url{https://gpt.space/blog/how-to-use-openai-model-temperature-for-better-ai-chat-responses)}}. Another important parameter is the \textit{maximum number of tokens} (\texttt{max\_tokens}), which determines the maximum length of the generated response. We set this to 3,019 tokens, which is the longest code snippet from the accepted answers on Stack Overflow. This value was chosen to ensure that the model-generated responses had sufficient scope to be complete while replicating the format and detail level found in the code snippets of our dataset.

Finally, because the focus of our research is on the use of libraries in large language models, we use regular expressions to extract the names of the used libraries from import statements in the model-generated code snippets. Additionally, we only extract the parent libraries and not the individual modules because our study's primary interest lies in understanding the usage of libraries. For example, from the import statement \texttt{from X import Y}, we extract \texttt{X}. This yields a total of 11,136 libraries (764 unique) across 10,000 question-code pairs.

%% file: results.tex
\section{Results}
\label{sec:results}
In this section, we present the results of our three research questions.
For each research question, we present our motivation, the approach
to answer the question, and key findings.

\input{rq1}

\input{rq2}

%% file: rq1.tex
\subsection*{RQ1: What are the characteristics of the software libraries recommended by ChatGPT?}

\noindent\textbf{Motivation.} 
As developers increasingly turn to tools like ChatGPT and other similar tools for coding assistance~\cite{hassani2023role}, the libraries included in the recommended code become more important as the code's performance, security, and functionality are directly impacted by these libraries~\cite{Latendresse_ASE2022, jafari_update, cox2019surviving}. Thus, in this RQ, we aim to understand the characteristics of libraries used in model-generated code to understand how software development is changing with the inclusion of LLMs. This knowledge will help us gain insights into the model's underlying knowledge base and its alignment with current software development practices. This is particularly relevant for developers relying on ChatGPT's suggestions to expedite their coding process. For that purpose, we will investigate the types of libraries recommended by ChatGPT, their popularity and maintenance characteristics, as well as their licenses. 




\noindent\textbf{Approach.} To answer this RQ, our approach involves a three-step analysis. First, we classify libraries into three categories: \textit{standard} (included in the standard Python package), \textit{third-party} (available on the PyPi repository), and \textit{other} (neither standard nor third-party). Subsequently, we examine the characteristics of libraries in ChatGPT-generated code. Finally, we analyze the licenses of third-party libraries. We outline each step below. 



First, To identify \textit{standard} Python libraries, we use \texttt{stdlib-list}~\footnote{\url{https://pypi.org/project/stdlib-list/}}, a Python package that lists all the standard libraries included in each Python version since 2.6. Then, for each library, we check if it is in the list of standard Python libraries. If so, we classify the library as \textit{standard}. For the libraries that were not classified as \textit{standard}, we verify if they are third-party by utilizing the PyPi API~\footnote{\url{https://warehouse.pypa.io/api-reference/index.html}}. If the API response returned a status code of 200, indicating a successful retrieval, we classify the library as \textit{third-party}. 

Second, we analyze the characteristics of third-party libraries. The reason we focus on third-party libraries is that standard libraries are always available with a Python installation, which means that characteristics such as the number of dependents and the number of dependencies do not apply in this context. We select the characteristics based on the work of Jafari et al.~\cite{jafari_update} on dependency update strategies and library characteristics, and the work of Vargas et al.~\cite{vargas_selecting} on practitioners' perspective on selecting third-party libraries. To obtain such characteristics, we utilize the \texttt{libraries.io} API, a tool that tracks open source software libraries across many programming languages and platforms. 

The third step is to identify the license category for each third-party library. For this, we manually labeled each one with the Software Package Data Exchange (SPDX) license identifier by examining the license information available on the PyPi page. For example, the library \texttt{numpy} has the SPDX identifier "BSD" which indicates that the license is a BSD style license. Then, we cross-referenced this information with the Open Source Initiative's (OSI)\footnote{\url{https://opensource.org/licenses}} list of approved licenses to identify the license category. For example, the BSD style of license belongs to the "permissive" license category. 

Finally, based on prior work on dependency analysis~\cite{jafari_update}, we group the library characteristics into three main categories:

\begin{itemize}
    \item \textbf{Popularity}: This category measures how widely adopted and trusted a library is within the Python ecosystem. We analyze three key characteristics to gauge popularity:
    \begin{itemize}
        \item Forks: The number of GitHub repositories that share the code and visibility settings with the original upstream repository of the library~\footnote{\url{https://docs.github.com/en/pull-requests/collaborating-with-pull-requests/working-with-forks/about-forks}}. 
        \item Stars: The number of GitHub users that have marked the library's repository as a favorite.
        \item Dependent Count: Indicates how many libraries rely on a specific library.
    \end{itemize}
    \item \textbf{Maintenance}: This category assesses the health, maturity, and activity level of the libraries based on the following characteristics:
    \begin{itemize}
        \item Dependency Count: Represents how many other libraries a particular library depends on. It is based on the library's state just before September 2021, which marks the date of the last data included in GPT-3.5 Turbo's training. 
        \item Age: The age of the project in months; calculated from when the library was first released until the date of metadata collection (February 2024).
        \item Version Frequency: The number of released versions divided by the age in months.
        \item SourceRank: The SourceRank metric of a library provided by libraries.io; derived from the latest version of a library as of the date of the metadata collection (February 2024).
    \end{itemize}
    \item \textbf{Licensing}: Licensing defines the terms under which the library can be used, modified, and distributed, which impacts the legal aspects of a software. In this category, we evaluate whether the recommended third-party libraries have an explicit usage license (indicated in the PyPi page of the library) and, if so, the category of the license (i.e., permissive, copyleft, weak copyleft, public domain). This categorization is essential to understanding the restrictions of using each library, which helps us determine their usage for different development scenarios. 
\end{itemize}

\noindent\textbf{Results.} Table~\ref{tab:libtypes} shows the distribution of the types of libraries found in ChatGPT-generated code against human code. The first column (Source) represents the source of the library (i.e., ChatGPT or Human). The second and third columns show the number of occurrences (\# Occ.) and the percentage of occurrences (\% Occ.) for each type of library, respectively. Finally, the last two columns show the number (\# Unique) and percentage (\% Unique) of unique libraries in each type, respectively. The results indicate that in 48.3\% of the cases, ChatGPT utilizes third-party libraries, compared to 45.2\% where standard Python libraries are used. In comparison, human-generated code from accepted answer snippets use third-party libraries in 38.7\% of the cases, highlighting a difference of 9.6\%. This suggests that ChatGPT includes more third-party libraries, especially when compared to human-generated. In 6.5\% of the cases, ChatGPT recommends libraries that fall into the \textit{other} category, indicating these libraries are neither identified as \textit{standard} nor \textit{third-party}. We delve deeper into this category and its implications in RQ2.


Table~\ref{tab:libcharacteristics} presents a statistical summary of the characteristics across the 437 third-party libraries used by ChatGPT. 

\textbf{Popularity.} In terms of popularity, the high standard deviation in the number of stars (10973.4), fork (2916.2), and dependents (5072.5) indicate a wide range in the popularity of the libraries used by ChatGPT. The maximum values observed (30,753 forks, 125,980 stars, and 58,040 dependents) suggest that some of these are highly popular, and are integral to a large number of projects (e.g., \texttt{requests}). In comparison, libraries found in human-generated code show lower mean values for forks (487.7) and stars (2,113.4), a similarly high maximum, indicating that while human developers also use popular libraries, ChatGPT does so more often. The median values further illustrate this, with ChatGPT having medians of 98 forks, 430 stars, and 98 dependents, compared to humans having medians of 7 forks, 23.5 stars, and 14 dependents. This suggest that for the same question and context, ChatGPT tends to favour more popular libraries than human developers.

\textbf{Maintenance.} Regarding maintenance, Table~\ref{tab:libcharacteristics} shows that the maximum number of dependencies for a library is 59 (e.g., \texttt{Zope}), which shows that some libraries are dependent on a high number of other libraries. However, the median value of 0 for dependency count indicates that a significant number of libraries are standalone or have minimal dependencies (e.g., \texttt{numpy} with 0 dependency or \texttt{scipy} with one dependency). This shows that ChatGPT has a tendency towards using libraries that do not require additional installations or configurations, which can potentially simplify the development process for users since the dependency count can impact the ease of integration, maintenance, and potential issues related to version conflicts or transitive dependencies~\cite{jafari_update, Latendresse_ASE2022}. In comparison, libraries found in human-generated code also show a median of 0 dependencies but with a slightly lower mean of 2.3. This shows a similar tendency towards human developers using libraries with fewer dependencies. 

Moreover, our results show that the overall SourceRank of the libraries is relatively high. According to the libraries.io documentation~\footnote{\url{https://docs.libraries.io/overview.html\#sourcerank}}, the maximum score for SourceRank is around 30 points, a difference of 8 points from the median SourceRank of 22 for the libraries used by ChatGPT. This suggests that more than half of the libraries are rated above the mid-range in terms of their overall quality and community trust and have a score close to the maximum possible score. Three libraries are found to be above this threshold: \texttt{requests}, \texttt{pandas}, and \texttt{numpy}, with SourceRank of 32 and 31, respectively. On the other hand, human developers use libraries with a median SourceRank of 12 and a mean of 13.3. This indicates that ChatGPT tends to use libraries that are recognized for their quality and maintenance more often than human developers.

Finally, we consider the age and version frequency of a library as interrelated characteristics because they provide insights into its maturity and maintenance activity~\cite{jafari_update}. The median age of the libraries used by ChatGPT is 159 months (approximately 13 years), while for libraries in human-generated code, it is around 125.1 months (approximately 10.5 years). This indicates in both cases, half of the libraries have been around for over a decade, implying a high level of maturity. The version frequency further illustrates the maintenance activity, with ChatGPT showing a median of 0.24 and a mean of 0.94, compared to human developers with a median of 0.16 and a mean of 0.59. This suggests that libaries used by ChatGPT are updated slightly more frequently than those used by human developers. 


\textbf{Licensing.}
Table~\ref{tab:licenses} shows the categories of licenses associated with the libraries used by ChatGPT compared to those used by human developers. Our results indicate a strong dominance of \textit{permissive} licenses among libraries used both by ChatGPT and human developers. Specifically, 71.7\% of the libraries used by ChatGPT fall under the permissive category, which includes licenses such as MIT, Apache 2.0, and BSD. This is only slightly higher than the 69.3\% observed in libraries found in the human-generated code from our dataset. This indicates that the vast majority of the libraries in both ChatGPT and human-generate code impose minimal restrictions on how they can be used, modified, and distributed. 

\begin{table}
    \centering
    \caption{Types of Libraries Used by ChatGPT.}

\input{tables/libtypes}
    \label{tab:libtypes}
\end{table}

In contrast, \textit{copyleft} licenses (e.g., GPL, LGPL, AGLP), which require derivative works to also be open source and distributed under the same license, are less prevalent. ChatGPT used libraries with copyleft licenses 14.2\% of the time, followed closely by 13.9\% in human-generated code. \textit{Public domain} licenses (e.g., CC0, Unlicense), which place the library in the public domain and relinquish all rights, have a low percentage of usage among both sets of libraries. ChatGPT includes public domain licenses in 1.5\% of cases, compared to 1.9\% in human-generated code. \textit{Weak copyleft} licenses (e.g., EPL, MPL), which are less restrictive than full copyleft but still require some openness for derivative works, represent 0.9\% and 1.1\% of the licenses in ChatGPT and human libraries, respectively. These results suggest that both ChatGPT and human developers might favor licenses that are either fully permissive or fully copyleft over licenses with more nuanced requirements. 

Interestingly, we observe the presence of libraries with unspecified licenses, which account for 10.4\% of the libraries used by ChatGPT and 12.9\% of the libraries used by human developers. The slightly higher percentage of unspecified licenses in libraries used in human-generated code could indicate that developers tend to overlook license specifications more often than ChatGPT. It could also suggest that human developers tend to include more obscure libraries where licensing information is not readily available (e.g., \texttt{app2}~\footnote{\url{https://pypi.org/project/app2/}}). The presence of unspecified licenses can be a potential risk for developers since it can potentially lead to legal ambiguities~\cite{largescalelicense2024}. 

Lastly, the "Other" category, which includes licenses that are less common or too obscure to classify (i.e., \textit{Other\/Proprietary}, \textit{Proprietary}, and \textit{Non-standard}), shows a minimal difference between libraries used by ChatGPT and human developers, at 1.3\% and 0.7\% respectively. These results indicate that in both cases, less conventional licensing arrangements are uncommon.

\begin{table*}
    \centering
    \caption{Characteristics of Libraries Used by ChatGPT.}
    \input{tables/libcharacteristics}
    \label{tab:libcharacteristics}
\end{table*}

\begin{table*}
    \centering
    \caption{Library License Categories.}
    \input{tables/licenses}
    \label{tab:licenses}
\end{table*}



\begin{center}\setlength{\fboxsep}{10pt}
\setlength{\fboxrule}{1pt}
\fcolorbox{gray!60}{gray!20}{%
    \parbox{0.9\columnwidth}{%
    Our findings suggest that ChatGPT tends to use libraries that are well-established and stable as opposed to popular according to their SourceRank, Age, and Version Frequency. We also found that ChatGPT tends to use libraries with fewer dependencies more frequently than those with a high dependency count, implying a preference for more standalone libraries. Moreover, similar to human developers ChatGPT tends to use libraries that have permissive licenses, with a smaller but still significant amount of libraries with copyleft and public domain licenses.}  
    }
\end{center}

%% file: tables/libtypes.tex
\begin{tabular}{llcccc}
    \toprule
   & \textbf{Source} & \textbf{\# Occ.} & \textbf{\% Occ.} & \textbf{\# Unique} & \textbf{\% Unique} \\
    \midrule
    \grayrow \textbf{Third-party} & ChatGPT & 5,377 & 48.3\% & 437 & 57.2\% \\
    & Human & 5,891 & 38.7\% & 720 & 57.8\% \\
    \grayrow \textbf{Standard} & ChatGPT & 5,029 & 45.2\% & 145 & 19\% \\
    & Human & 7,538 & 49.6\% & 159 & 12.8\% \\
    \grayrow \textbf{Other} & ChatGPT & 730 & 6.5\% & 182 & 23.8\% \\
    & Human & 1,778 & 11.7\% & 366 & 29.4\% \\
    \midrule
    \grayrow \textbf{Total} & ChatGPT & 11,136 & 100\% & 764 & 100\% \\
    & Human & 15,207 & 100\% & 1,245 & 100\% \\
    \bottomrule
\end{tabular}

%% file: tables/libcharacteristics.tex
\begin{tabular}{@{}lllrrrrrr@{}}
\toprule
\textbf{Category} & \textbf{Characteristic} & \textbf{Source} & \textbf{Median} & \textbf{Mean} & \textbf{Std. Deviation} & \textbf{Min} & \textbf{Max} \\ \midrule
\grayrow \textbf{Popularity} & No. of forks & ChatGPT & \textbf{98} & 907.6 & 2916.2 & 0 & 30753 \\
         &       & Human  & \textbf{7}  & 487.7  & 2063.4   & 0 & 30497 \\
\grayrow & Stars & ChatGPT & \textbf{430} & 3866 & 10973.4 & 0 & 125980 \\
         &       & Human  & \textbf{23.5 } & 2113.4  & 7796.5   & 0 & 127601 \\
\grayrow & No. dependents & ChatGPT & \textbf{98}  & 1328.5 & 5072.5 & 0 & 58040 \\
         &       & Human  & \textbf{14}  & 840.8  & 4205.4   & 0 & 60232 \\
\midrule
\grayrow \textbf{Maintenance} & No. dependencies & ChatGPT & \textbf{0}  & 3.2 & 7.5 & 0 & 59 \\
         &         & Human  & \textbf{0}  & 2.3  & 6  & 0 & 59 \\
\grayrow & sourceRank & ChatGPT & \textbf{22}  & 15.9 & 7 & 1 & 32 \\
         &         & Human  & \textbf{12}  & 13.3 & 7 & 0 & 32 \\
\grayrow & Age (months) & ChatGPT & \textbf{159} & 116.9 & 59.7 & 2.6 & 259 \\
         &        & Human  & \textbf{125.1}  & 117.2  & 58.6  & 4.3 & 261 \\
\grayrow & Version frequency & ChatGPT & \textbf{0.24} & 0.94 & 2.4 & 0.01 & 21.2 \\
         &         & Human  & \textbf{0.16}  & 0.59  & 1.6  & 0.01  & 15.5 \\
\bottomrule
\end{tabular}


%% file: tables/licenses.tex
\begin{tabular}{llcc|cc}
    \toprule
    \multicolumn{2}{l}{} & \multicolumn{2}{c|}{\textbf{ChatGPT}} & \multicolumn{2}{c}{\textbf{Human}} \\
    \cmidrule(lr){3-4} \cmidrule(lr){5-6}
    \textbf{Category} & \textbf{Definition} & \textbf{\# Occ.} & \textbf{\% Occ.} & \textbf{\# Occ.} & \textbf{\% Occ.} \\
    \midrule
   \grayrow \textbf{Permissive} & Allows broad reuse with  & 324 & 71.7\% & 500 & 69.3\% \\
   \grayrow  & minimal restrictions. &  &  &  &  \\
    \textbf{Copyleft} & Requires derivative works  & 64 & 14.2\% & 100 & 13.9\% \\
    & to use the same license. &  &  &  &  \\
   \grayrow \textbf{Public Domain} & No exclusive intellectual  & 7 & 1.5\% & 14 & 1.9\% \\
   \grayrow & property rights. &  &  &  &  \\
    \textbf{Weak Copyleft} & Allows linking with other licenses & 4 & 0.9\% & 8 & 1.1\% \\
    \grayrow \textbf{Other} & Miscellaneous licenses not in the & 6 & 1.3\% & 5 & 0.7\% \\
   \grayrow & above categories such as "Proprietary", &  &  &  &  \\
    \grayrow & "Non-standard", and "Other\/Proprietary". &  &  &  &  \\
   \textbf{Unspecified} & License not specified. & 47 & 10.4\% & 93 & 12.9\% \\
    \midrule
    \textbf{Total} & & 452 & 100\% & 722 & 100\% \\
    \bottomrule
\end{tabular}

%% file: rq2.tex
\subsection*{RQ2: What are the challenges encountered by developers when using ChatGPT for library recommendations?}

\noindent\textbf{Motivation.} 
In RQ1, we find that certain libraries recommended by ChatGPT are neither part of the standard Python libraries, nor published on the PyPi repository. Such libraries are potentially problematic because it means that they do not work out of the box, which can confuse developers and hinder their workflow. Thus, identifying and understanding challenges such as this is crucial for improving the practical utility of LLMs in real-world settings. When ChatGPT uses libraries that do not work out of the box (i.e., they do not exist on the current PyPi repository and are not part of the default Python installation), it not only impacts the productivity of developers but also raises concerns about the trustworthiness and validity of the generated code~\cite{adamson2023assessing}. By studying the challenges developers encounter in the context of LLMs as software librarians, we can develop actionable strategies to mitigate them. Thus, this RQ aims to pinpoint the specific difficulties developers face when using ChatGPT for library recommendations. 


\noindent\textbf{Approach.}
To understand the underlying causes behind the presence of problematic libraries in ChatGPT-generated code, we first determine whether these libraries were explicitly mentioned in the question. If so, we categorize the library as \textit{hard-coded}, indicating that ChatGPT was directed to use a library not found in standard Python libraries or the PyPi~\cite{pypi} repository. 

In cases where the unclassified library is not hard-coded, we verify whether it is deprecated (as of March 2024). We devised a two-step approach to determine the deprecation status of a library. The first step targets libraries that while not found on the PyPi registry, might have been previously part of the standard Python installation, and involves querying the Python documentation website~\cite{python-docs} for Python 3 and Python 2 versions. In this case, we consider a library deprecated if the request for Python 3 is successful (i.e., status code is 200) but with a changed library name in the response URL (e.g., \texttt{urllib2} becomes \texttt{urllib.request}). Alternatively, if the request for Python 3.x fails (e.g., status code is 404), we request the documentation website for Python 2.x; a successful request leads to the library being labeled as deprecated.

The second step covers the libraries that might have been on the PyPi registry, but have been removed since then. This step involves using virtual environments and the \texttt{subprocess} module to automate the process. For each library, we create a unique virtual environment and install the library. During both the installation and import stages, we check for deprecation warnings. This is done by capturing and analyzing the output from the \texttt{pip install} command and a Python script that imports the library. 


Next, once we have ruled out the possibility of a problematic library being deprecated or hard-coded, we look at other possibilities. For this, we take the remaining set of recommended libraries and perform open coding. This process involves analyzing the Stack Overflow question used to generate the prompt, including the question body, any provided code snippets, and the accepted answer's body. Through this analysis, we developed a set of nine mutually exclusive labels to categorize the libraries. The open coding was performed by one author, with the labeling scheme having been collaboratively developed by both authors. After the initial labeling, the results were validated through discussions with the second author. The reason behind our decision to involve only one author in the labeling process is that the nature of the labeling in this context is largely factual rather than subjective. For example, \texttt{cv2} is an alias for \texttt{opencv-python}, is a factual classification based on established naming conventions and not open to interpretation. Thus, the labels in our scheme were defined based on concrete characteristics of the libraries and their usage.

\noindent\textbf{Results.} Table~\ref{tab:inaccurate_lib_types} shows the number and ratio of the problematic library recommendations made by ChatGPT. Below we outline each of the categories that resulted in an import or installation failure. 

\textbf{Hard-coded Libraries (32.9\%).} A significant portion (one-third) of problematic recommendations are hard-coded libraries. This implies that ChatGPT generates responses based on contextually mentioned libraries, regardless of their existence or current support status. Such behavior can mislead developers into believing that the requested library is real, even if it doesn't exist, undermining the trustworthiness of LLM-generated code.

\textbf{Alias (30.3\%).} Another prevalent challenge arises from ChatGPT's use of aliases in library recommendations. The Python ecosystem lacks a standardized naming convention for distributions, leading to discrepancies between library import names and installation names~\cite{revisit-dist-name}. For example, the library \texttt{opencv} has the distribution name \texttt{opencv} but is imported as \texttt{cv2}. While aliases themselves aren't inherently problematic, they introduce a layer of complexity that developers need to navigate. This highlights the need for LLMs to be aware of these naming conventions and favor explicit imports (e.g., \texttt{from numpy import array}) to enhance code clarity and avoid potential errors.

\textbf{Module (16.2\%).} Our results reveal challenges related to how ChatGPT distinguishes between modules and libraries. In Python, a library is a collection of modules, while a module is a single Python file. Python provides several ways to import modules and libraries, which we categorize into "implicit" and "explicit" imports. LLMs should understand these nuances and adapt their recommendations accordingly. Inappropriate import statements (e.g., importing \texttt{array} without specifying the need for \texttt{numpy}) can lead to confusion and installation failures if developers attempt to install the module as a standalone entity.

\textbf{Placeholder (14.7\%).} Placeholders appear as generic or illustrative names that possibly intend to guide developers, and are often used when the question is more vague (e.g., Pythonic way to write long import statements~\footnote{\url{https://stackoverflow.com/questions/26253545/pythonic-way-to-write-long-import-statements}}). Placeholders can become problematic when what is intended to be a generic example ends up being interpreted as an actual library. For instance, in our dataset, there was a case where the import statement was \texttt{import ParentModel}, which was intended as a placeholder. This ambiguity can cause confusion, where it would be more beneficial for ChatGPT to provide actionable examples as oppose to illustrative examples. 

\textbf{Deprecated (1.4\%).} A smaller but still significant portion of problematic libraries are tied to deprecated libraries. Such libraries may have outdated functionality, or are no longer supported or available through PyPi. A particular case that may generate confusion is when the syntax of the library is outdated. For instance, the \texttt{queue} library is imported as \texttt{Queue} in Python 2, but is imported as \texttt{queue} in Python 3. This suggests that models should have updated, timely knowledge of library statuses.

\textbf{Mistake (1\%).} The \textit{Mistake} category represents cases where ChatGPT appears to have made genuine errors in suggesting libraries. Table~\ref{tab:mistakes} shows specific instances where ChatGPT recommended non-existent libraries or inappropriate import statements in response to various StackOverflow questions from our dataset. For example, the suggestion \texttt{fuzzyhashmap} in response to the question about fuzzy hash tables in Python is not a legitimate library and indicates that ChatGPT might have fabricated the name based on related terminology in the question. 

\textbf{Other (1.2\%).} The \textit{Other} category groups recommendations for custom libraries (libraries not widely available, e.g., \texttt{progress\_statusbar} refers to a custom module from a Python development cookbook~\cite{precord2015wxpython}), external applications that can be extended with Python (but not installed via PyPi), and libraries dependent on specific environments (e.g., Android Python interpreter).

\begin{table}
    \centering
    \caption{Causes for Import or Installation Failures of ChatGPT Library Recommendations.}
    \label{tab:inaccurate_lib_types}
    \input{tables/inaccurate_lib_types}
\end{table}

\begin{table*}
    \centering
    \caption{Incorrect library imports by ChatGPT based on Stack Overflow questions.}
    \input{tables/mistakes}
    \label{tab:mistakes}
\end{table*}

\begin{center}\setlength{\fboxsep}{10pt}
\setlength{\fboxrule}{1pt}
\fcolorbox{gray!60}{gray!20}{%
    \parbox{0.9\columnwidth}{%
    Our findings indicate that two thirds of problematic libraries are hard-coded or alias-driven, with 32.9\% and 30.3\%, respectively. Our analysis also reveals a mix-up between modules and libraries, the introduction of placeholders, as well as context-specific recommendations. Such library recommendations can lead developers into confusion and waste of time by trying to unsuccessfully import or install the recommended library.
        }
    }
\end{center}

%% file: tables/inaccurate_lib_types.tex
\begin{tabular}{lccc}
    \toprule
    \textbf{Cause} & \textbf{\# Occ.} & \textbf{\% Occ.}\\
    \midrule
    \textbf{Hard-coded} & 241 & 33\% \\
    \textbf{Alias} & 223 & 30.5\% \\
    \textbf{Module}  & 119 & 16.3\% \\
    \textbf{Placeholder} & 108 & 14.8\% \\
    \textbf{Deprecated} & 10 & 1.4\% \\
    \textbf{Mistake}  & 7 & 1\% \\
    \textbf{Other} & 22 & 3\% \\
    \midrule
    \textbf{Total} & 730 & 100\% \\
    \bottomrule
\end{tabular}

%% file: tables/mistakes.tex
\begin{tabular}{ll}
    \toprule
    \textbf{StackOverflow Question} & \textbf{Import Statement} \\
    \midrule
    \grayrow
    "Fuzzy" hashtables in Python to replace elif chains? & \texttt{import fuzzyhashmap} \\
    Can I work with a socket I created in another file in python? & \texttt{import create\_socket} \\
    \grayrow
    How can I use ptiPython with Flask-Script together? & \texttt{import flask\_ptipython} \\
    Working with Import Statements in OSX Mavericks. & \texttt{import opencv},  \texttt{import scikit},\\
    & \texttt{import Queue} \\
    \grayrow
    Java and Python implementation of Blowfish produce- & \texttt{import javax} \\
    \grayrow
    different results. & \\
    \bottomrule
\end{tabular}

%% file: discussion.tex
\section{Discussion}
\label{sec:discussion}
In this section, we discuss the implications of our work and propose recommendations for practitioners and researchers in the context of LLMs as software librarians. 

\subsection{Are LLMs Good Software Librarians After All?}
\label{sec:good_librarians}
Our results indicate that ChatGPT generated only five hallucinations out of 10,000 questions, and only 6.5\% of the libraries were potentially problematic. Based on these findings, we argue that LLMs like ChatGPT have the potential to be good software librarians, though several important considerations must be addressed, which we discuss below. 

\noindent\textbf{Licensing Awareness.} One significant area of improvement for ChatGPT as a software librarian is its lack of communication of software licenses. While most of the recommended libraries had permissive licenses, 14.2\% were copyleft licensed- a more restrictive type of license- and 10.4\% had no license specified. This was not explicitly communicated by the model, which can lead to potential legal and compliance issues for developers. Permissive licenses, such as the MIT or Apache licenses, allow developers to use, modify, and distribute the software with minimal restrictions, providing flexibility and reducing legal risks. On the other hand, copyleft licenses, such as the GNU General Public License (GPL), require that any derivative works also be distributed under the same license. This implies that if developers use a GPL-licenses library in their project, they may be obligated to release their own code under the GPL as well. This can be problematic for proprietary software developers who do not wish to open-source their code. To better understand this, we can consider a scenario where a developer is working on a proprietary project and uses ChatGPT to generate code. The model recommends a library without specifying its license, for example \texttt{PyQT~\footnote{\url{https://pypi.org/project/PyQt5/}}}, one of the most popular cross-platform GUI libraries for Python. The developer, assuming it is safe to use, integrates the library into their codebase, but later discovers that it is GPL-licensed. This forces the developer to either comply with GPL requirements, which may not be feasible or desirable, or replace the library, which involves additional effort to refactor the code and ensure compatibility.

To avoid such problematic scenarios, developers can enhance the model's responses by including safeguards such as explicit license information for each recommended library. For example, when suggesting a library, the model should include a note such as "This library is licensed under the GNU GPL v3.0". Moreover, if the generated output contains a library, the response should clearly state the license type and restrictions, e.g., "The library \texttt{PyQt5} is licensed under the GNU GPL v3.0, a copyleft license that may require your code to also be open-sourced under the same license."

Also, LLM (including GPT) developers should train models to detect and flag libraries with restrictive licenses, and provide alternatives with permissive licenses where possible. For instance, if the model detects a GPL-licensed library, it should offer an alternative like: "While \texttt{PyQt5} is suitable, it is GPL-licensed. You might consider \texttt{Tkinter}, which has a more permissive license (PSF License Agreement)."


\noindent\textbf{ChatGPT Unconditionally Recommends Libraries.} One notable finding from RQ2 is that 33\% of the libraries that did not work out of the box in ChatGPT code were actually explicitly mentioned in the user question. This highlights a critical issue: if a user explicitly instructs the model to use libraries that are not standard, recognized, or even existent, ChatGPT will still attempt to generate code utilizing that library without any warning. This can lead to several significant problems. 

First, ChatGPT may generate code for libraries that are non-existent or no longer supported. This can lead to confusion and wasted time as developers attempt to install and use a library that simply isn't there. For example, a user asked "how to post data and binary data using urllib2 in Python?"~\footnote{\url{https://stackoverflow.com/questions/36479006/how-to-post-data-and-binary-data-using-urllib2-in-python}}. In this case, \texttt{urllib2} is a library that is no longer supported and no longer available on PyPi. Despite this, ChatGPT generated code using \texttt{urllib2} as shown in Listing~\ref{list:urllib2ex}. While the code itself is correct, installing \texttt{urllib2} is not possible with Python 3. Although the Stack Overflow question from which the prompt was generated is eight years old, tools like ChatGPT should have up-to-date knowledge of libraries to ensure that developers use current and supported libraries in their code. From the developer's side, fixing such issues requires the developer to recognize that the library does not exist or is deprecated, which may not be immediately obvious, especially to less experienced developers. Thus, the reliance on user knowledge to identify and correct library recommendations undermines the efficacy and reliability of LLMs as software librarians. As such, we propose that LLM developers enhance models to include warnings when a user requests a non-existent or deprecated library along with the generated code. For example, if a user asks for a deprecated library like \texttt{urllib2}, the model should respond with, "The library \texttt{urllib2} is no longer supported. Consider using \texttt{requests} instead." This will help developers avoid wasting time on outdated libraries and ensure they use current, supported ones.


Furthermore, there is a critical security implication arises when ChatGPT unconditionally generates code for libraries that are explicit in the request. For example, if a user mistakenly asks for \texttt{padnas} instead of \texttt{pandas}, ChatGPT might generate code for the misspelled library name. If the user attempts to install \texttt{padnas}, they might inadvertently install a malicious library that takes advantage of typosquatting- a common tactic where attackers upload malicious packages with names similar to popular libraries to exploit such mistakes. In this specific example, \texttt{padnas}~\footnote{\url{https://pypi.org/project/padnas/}} is a harmless library designed to resemble \texttt{pandas} to prevent such attacks, which highlights that these scenarios are not uncommon.


To mitigate these security risks, we recommend that both LLM developers and practitioners take proactive measures. For example, LLM developers should enhance models to verify and correct library names before generating code. If the model detects a likely typo, it should suggest the correct library name, such as, "It looks like you meant \texttt{pandas}. Generating code for \texttt{pandas} instead of \texttt{padnas}". Practitioners should also adopt best practices when using LLMs such as ChatGPT to assist in programming tasks~\cite{marvin2023prompt}. Notably, they should cross-check library names and ensure they are correctly spelled and widely recognized before attempting to install them. It is critical that practitioners do not blindly adopt generated code without first verifying the overall health of the libraries, as the health of libraries is constantly evolving, and new vulnerabilities may have emerged since ChatGPT was last trained. For this, practitioners can use tools such as the sourceRank provided by libraries.io, or Snyk~\footnote{\url{https://snyk.io/}}, a security platform for securing code, open source dependencies, and cloud platforms. 

\begin{lstlisting}[language=Python,caption={ChatGPT-generated code for deprecated urllib2.},captionpos=b,label=list:urllib2ex]
    import urllib2
    import json
    
    url = 'http://url.com?u=user&p=pass'
    data = json.dumps({'config':'configData'}) 
    cont_len = len(data)
    req = urllib2.Request(url, data, 
        {'Content-Type': 'application/json', 'Content-Length': cont_len})
    
    f = urllib2.urlopen(req)
    response = f.read()
    f.close()
\end{lstlisting}

We conclude that while LLMs like ChatGPT can be good software librarians, they can be improved to better assist developers in streamlining their work. This can be achieved by implementing warning mechanisms to alert users when they request libraries that are deprecated or potentially non-existent, integrating verification tools that cross-reference libraries against known repositories like PyPi, and incorporating alerts for potential security risks like typosquatting. Additionally, providing more explicit information and context for recommended libraries, such as licensing, will help users mitigate potential conflicts and ensure easier integration into their codebase. 

\subsection{The Impact on Dependency Management}
\label{sec:dependency_management}
Our study highlights several implications for practitioners who are increasingly relying on LLMs as programming assistants. Here, we discuss the potential impact on dependency management and development workflow. 

\noindent\textbf{Increased Use of Third-Party Libraries.} Our results show that ChatGPT tends to use third-party libraries more frequently than human developers, favoring well-established and widely adopted libraries. The use of third-party libraries can significantly impact dependency management as it introduces additional complexity and risks into software projects~\cite{wang2020empirical, raemaekers2011exploring}. Each new library brings its own set of dependencies, the majority of which are transitive (dependencies induced by other dependencies)~\cite{Latendresse_ASE2022}. This can lead to \textit{dependency hell}, a term used to describe when there are so many dependencies and potential version conflicts that resolving them becomes a significant challenge and developers spend more time managing dependencies than writing actual code~\cite{mujahid2021effective, chen2021helping}. 

To address these challenges, LLM developers should enhance models to provide not only code with appropriate libraries but also information on the dependency footprint of each library. For example, when suggesting a library, the model could include a note such as "This library has 5 direct dependencies and 20 transitive dependencies." Moreover, LLMs could be trained to recommend libraries with fewer dependencies when multiple options are available to minimize the risk of dependency hell. LLM developers should also incorporate security and stability checks into the models by leveraging databases like the National Vulnerability Database~\footnote{\url{https://nvd.nist.gov/}} (NVD) or GitHub Security Advisories~\footnote{\url{https://github.com/advisories}} (GSA).

In addition to the steps for LLM developers, we recommend that practitioners thoroughly evaluate the maintenance metrics of libraries recommended by ChatGPT before integrating them into their codebase. Firstly, practitioners should consider the number of dependencies associated with each library. A high number of dependencies increases the risk of conflicts and broadens the attack surface of the codebase~\cite{wang2020empirical, Latendresse_ASE2022}. For instance, using tools like \texttt{npm ls} for JavaScript or \texttt{pipdeptree} for Python can help visualize and asses the dependency tree of a library. Practitioners should also consider the version frequency of a library; while frequent updates may indicate active maintenance, they can also be a sign of instability, and managing constant updates can be challenging~\cite{jafari_update}. Tools such as Dependabot~\footnote{\url{https://github.com/dependabot}} or Renovate~\footnote{\url{https://github.com/renovatebot/renovate}} can automate the process of dependency updates and help keep track of changes without overwhelming the development process.


\noindent\textbf{Libraries Recommended by ChatGPT are Frozen in Time.} The knowledge of an LLM is essentially "frozen in time" at the point when it was last trained. This arises from the nature of training LLMs, which is an expensive and infrequent process coupled with the dynamic nature of libraries, which continually evolve with updates with add new features, fix bugs, and address security vulnerabilities. As a result, the libraries recommended by ChatGPT may not reflect the most current versions. Thus, practitioners might find that the functions or methods found in ChatGPT-generated code are deprecated or have been replaced by more efficient alternatives in the latest library versions. New versions of libraries can include breaking changes that are not compatible with previous versions, which can cause runtime errors and necessitate additional effort to refactor the code with work with the updated version. Moreover, libraries are frequently updated to address security vulnerabilities~\cite{jafari_update, Latendresse_ASE2022}, and using an outdated snapshot of a library version as recommended by ChatGPT to avoid compatibility issues can expose projects to security risks. 

To mitigate these issues, we recommend that practitioners integrate Continuous Integration (CI) tools into their workflow to automate the testing of code against the latest versions of dependencies. CI tools can help identify compatibility issues early, allowing developers to address them before they become problematic~\cite{hilton2016usage}. Furthermore, LLM developers should enhance models by incorporating mechanisms to check the currency of recommended libraries. For instance, the LLM could provide a disclaimer such as, "This library is based on a version as of [last training date]. Please verify with the latest documentation." This would prompt practitioners to validate the suggested library version. LLM developers can also train models to prioritize stable libraries that have low frequency of breaking changes. For example, if multiple libraries provide similar functionalities, the model should recommend the one with a proven long-term stability, not necessarily the most popular one as our results of RQ1 indicate. This could be done by integrating data from sources like GitHub releases or library changelogs.

\noindent\textbf{Mind Your Prompts.} In light of the above-mentioned implications, one might consider asking ChatGPT for alternative standard libraries equivalent to third-party libraries. Standard libraries are generally more stable, better documented, and maintained as part of the core language distribution, which reduces the dependency management burden and minimizes the risks as mentioned above~\cite{edmunds-2023}. However, when asking ChatGPT for alternative libraries, practitioners should be mindful of the potential for the model to "hallucinate". An article by Vulcan Cyber states that between 25\% and 40\% of the libraries generated by ChatGPT are hallucinations~\cite{lanyado-2023}. In this article, the author asked ChatGPT to generate code to integrate a library in \texttt{Node.js} and then repeatedly requested alternative methods. As a result, ChatGPT began hallucinating and producing non-existent libraries. While this has significant security implications, it does not reflect a typical user interaction with ChatGPT, which we aimed to reflect in this study. 

To mitigate these risks, we recommend that practitioners craft solid prompts from the start, providing enough context surrounding the type of library they are seeking so that the model understands what is expected early on. For example, instead of asking, "Canyou suggest a library for HTTP requests in Python?", a more detailed prompt would be, "Can you suggest a standard Python library or widely-used third-party library for making HTTP requests, and provide a brief explanation of its advantages?" LLM developers can also help in addressing these issues by enabling the models to recognize when it is best to use a standard library versus a third-party library. Implementing a confidence scoring system for library recommendations, where the model indicates the certainty of its suggestions, can guide LLM users in evaluating the reliability of the recommendations. This can be integrated in the models' responses such as, "This library has a high confidence level based on currenty training data."


\noindent\textbf{Users as Active Evaluators.} The results of RQ2 suggest that the context in which a library is recommended is crucial in determining its relevance and utility. Thus, problematic library usage is not inherently tied to ChatGPT making mistakes, but to the very context-dependent nature of the PyPi ecosystem. This means that developers using LLMs as programming assistants are not just consumers of the library recommendations but should be active evaluators to ensure that the libraries they integrate into their own code are compatible with their own context. For the LLM development community, this means improving the models' ability to recognized, understand, and prioritize context-specific information when generating code with software libraries. For instance, the model should consider a project's framework, the programming language version, and the specific functionality required.







%

%% file: related_works.tex
\section{Related Works}
\label{sec:related_works}
In this section, we discuss the related literature divided into two aspects. First, we discuss the works that have focused on dependency management and related challenges. Second, we discuss the works that report on the use of large language models as programming assistants. 

\subsection{Dependency Management Challenges}
\label{sec:rec_systems}

While open source software libraries significantly reduce development time and costs~\cite{basili1996reuse, decan2019empirical, murphy2019predicts}, depending on numerous libraries introduces complexity and potential dependency management challenges~\cite{Latendresse_ASE2022, mujahid2021effective, wang2020empirical}. One such challenge is highlighted by Mujahid et al.~\cite{mujahid2023characteristics} who identify library selection as a critical aspect of dependency management. In their work, they surveyed developers from the npm ecosystem to qualitatively understand the characteristics of highly-selected libraries. Their results show that JavaScript developers believe that such libraries are well-documented, popular, and free of vulnerabilities. Building upon this work, our study leverages those same characteristics to categorize libraries used by ChatGPT.Hauge et al.~\cite{hauge2009empirical} show that organizational library selection is an ad-hoc process that often relies on a combination of past experiences, expert advice, and online resources. This is further discussed in the work of Haenni et al.~\cite{haenni2013categorizing} where the authors surveyed developers about their decision-making when selecting a library to integrate into their application. Their findings show that, in general, developers do not apply rationale when selecting libraries. Alternatively, developers opted for libraries that fulfilled the immediate task requirements. Given this lack of formal selection processes and the increasing popularity of LLMs as programming assistants, this motivated us to investigate the role of LLMs as software librarians. 

Dependency hell is a concept discussed in several studies~\cite{fan2020escaping, tanabe2018context, abdalkareem2020impact} and refers to when a project has an excessive number of dependencies, and managing these dependencies becomes difficult and error-prone. Chen et al.~\cite{chen2021helping} discuss the impact of "trivial packages," referring to libraries implementing simple functionalities, on the npm ecosystem. Their survey highlights that developers struggle with the multiple dependencies introduced by these libraries, contributing to dependency hell. For instance, a developer reported the cascading effect of patching a deeply nested dependency, requiring updates throughout the dependency tree. Jafari et al.~\cite{jafari_update} investigate the relationship between npm library characteristics and the dependency update strategy opted by its dependents. The authors report that the release status, the number of dependents, and the age of a library are the most important indicators of the dependency update strategy. Raemaekers et al.~\cite{raemaekers2011exploring} discuss the risks associated with the usage of third-party libraries. They identify key library attributes that could serve as risk indicators. Notably, they report that more popular libraries may be updated more frequently, which increases the chance for new bugs to get introduced into the codebase. These findings support our claim that LLMs can be improved software librarians by providing critical information about recommended libraries, such as maintenance metrics (e.g., number of dependencies, version update frequency, age). Such transparency can help developers anticipate and manage potential maintenance challenges associated with the increased complexity of dependencies. 

\subsection{LLMs in Software Development Workflows}
\label{sec:llms_dev_workflows}

LLMs are increasingly used by software engineering practitioners to perform various tasks, such as code generation, and have shown the potential to improve developer productivity~\cite{ross2023programmer, heitz2024evaluation, teubner2023welcome}. However, some studies have raised concerns about the reliability of LLM-generated code. For instance, Zhong et al.~\cite{zhong2024chatgpt} report common API misuse patterns found in popular LLMs. The study reveals that in the case of GPT-4, 62\% of the generated code contained API misuses. The authors argue that this is particularly problematic given that users of LLM code generation are generally not familiar with the APIs that LLMs generate code for, and cannot tell whether the provided code is correct or not. Similarly, a Vulcan Cyber article claims that ChatGPT hallucinated in almost 40\% of the programming questions it was asked~\cite{lanyado-2023}. Our findings diverge from those reported in the aforementioned studies. This discrepancy might be attributed to differences in experimental design. Zhong et al. employed "one-shot" or "few-shot" approaches, providing either irrelevant or relevant examples alongside the prompt. The Vulcan Cyber article describes a scenario where ChatGPT was instructed to generate code for a specific library and then repeatedly prompted for alternatives, potentially leading to hallucinated libraries. In contrast, our study aimed to simulate a more realistic developer-LLM interaction by prompting ChatGPT only once. This approach aligns with findings from Jin et al.~\cite{jin2024can} whose empirical study reveals that developers only request code regeneration from ChatGPT in 3\% of conversations. 

While LLMs are making significant advancements in software development, concerns regarding reliability and the potential for hallucinations remain. However, studies also highlight the potential of LLMs to improve developer productivity through functionalities like code completion and search. Ross et al.~\cite{ross2023programmer} developed an LLM-based programmer's assistant and evaluated their system on 42 participants. Their results reveal that participants were in majority positive in the assistant's potential for improving their productivity. Heitz et al.~\cite{heitz2024evaluation} evaluate and compare the performance of OpenAI's ChatGPT and Google's Gemini in programming code. They find that while the premium version of the models offers enhanced performance, their free counterparts remain highly relevant for a wide range of users in the context of software development. Also, the authors report that these models can significantly accelerate coding tasks and improve productivity, but necessitates rigorous, especially if the generated code is used in critical areas. Despite this, there is a gap in understanding how LLMs perform as software librarians, a critical role in the development process that impacts project maintainability, security, and compatibility. 

Our study directly addresses this gap by investigating the effectiveness of LLMs in recommending libraries. We analyze libraries suggested by ChatGPT in response to real-world developer prompts from Stack Overflow questions. This allows us to assess LLM performance in a context that mimics actual developer usage and avoids potential biases introduced by experimental setups. By evaluating factors like library popularity and maintenance, licensing, and potential dependency challenges, our work aims to inform the development of more robust LLM software librarians, and provide developers with recommendations and considerations when using such tools to streamline their workflows.

%% file: threats.tex
\section{Threats to Validity}
\label{sec:threats}

\noindent\textbf{Internal validity} considers the experimenter's bias and errors. Thus, a potential threat to the validity of our study is the manual categorization process of instances where a library import or installation resulted in a failure. To mitigate this, each case was discussed and reviewed by other authors and any inconsistencies were carefully addressed. Additionally, the parameters used when prompting the model can impact the generated responses. We addressed this by using ChatGPT's default parameters, mimicking a typical user interaction. Also, a time gap between the Stack Overflow questions and the generated code could introduce inconsistencies due to library changes over time. To mitigate this, we used the latest version of the libraries (as of February 2024) as a baseline for our analysis. Moreover, while the open coding process for categorizing the libraries was conducted by a single author, the labeling scheme was developed collaboratively and the results were validated through discussions with other authors. We opted not to use multiple coders or statistical measures like Cohen’s kappa due to the objective nature of the labeling, which focused on factual classifications (library characteristics and usage) rather than subjective interpretations. Nevertheless, this decision may introduce a potential bias, although it was minimized through careful validation and clear, objective criteria. Finally, the stochastic nature of LLMs might lead to variations in generated code for the same prompt. To enable the replicability of this study, we provide our dataset of generated code snippets. 

\noindent\textbf{External validity} considers the generalizability of
the findings. First, it is possible that the Stack Overflow code was part of ChatGPT's training data, which might influence its library recommendations by favoring libraries commonly used on that platform. To mitigate this, we compared the libraries used by ChatGPT to those used by human developers, allowing us to identify potential biases towards Stack Overflow-popular libraries. Next, our analysis is based on a sample of 10,000 Stack Overflow questions, which may not fully capture the diversity of real-world programming tasks and library usage patterns. To address this, we referred to the literature to identify typical developer interaction with ChatGPT to design our experimentation setup. Moreover, our focus on Python code might not translate to other programming languages. Despite the popularity of the Python language in software development, especially on Stack Overflow~\cite{stack-compare}, investigating additional languages would be valuable to understand if these models exhibit similar or different behaviors. Finally, we only investigated the performance of one specific LLM (GPT 3.5 Turbo). Including a wider range of LLMs in future studies would provide a more comprehensive picture of how these models handle library recommendations.

%% file: conclusion.tex
\section{Conclusion and Future Work}
\label{sec:conclusion}
This study investigated the effectiveness of LLMs as "software librarians" by analyzing libraries suggested by ChatGPT (specifically GPT-3.5 Turbo) for real-world coding problems derived from Stack Overflow questions. Our findings shed light on several key aspects of LLM library recommendations. First, ChatGPT uses third-party libraries 10\% more often than human developers, and prioritized stable and well-established options. Additionally, ChatGPT tends to recommend libraries with few dependencies, which suggests a focus on more self-contained solutions. Interestingly, the libraries used by ChatGPT have a majority of permissive licenses, but a concerning amount of copyleft licenses, indicating that developers need to exercise extra caution in the way they use and distribute the generated code. As such, we recommend that tools like ChatGPT provide explicit information about licensing, providing alternatives to libraries with more restritrive usage. 

Our analysis also revealed areas for improvement. We found that 6.5\% of the libraries in ChatGPT-generated code led to import or installation failures and stemmed from aliases and implicit imports of modules. Moreover, two thirds of the problematic libraries were explicitly mentioned in the user question, indicating that ChatGPT generates code for libraries that may not be recognized or even existent. These issues can lead to developer confusion and frustration when encountering import or installation failures. We recommend that developers practice caution before integrating libraries recommended by ChatGPT into their codebase by cross-referencing the libraries with up-to-date resources such as official documentation. We also recommend that LLM programming assistants provide explicit information maintainability metrics of recommended libraries to guide developers in selecting and integrating appropriate libraries into their codebase. In the future, we plan to implement a tool that integrates with LLMs to perform checks on libraries and provide essential metadata (e.g., the number of dependencies and the version frequency) and license information. Additionally, we plan to replicate our study to include more models such as CodeT5 and Llama3.

We conclude this work by answering the question: Is ChatGPT a good software librarian? Our answer is a cautious yes. 
